%% file: main.tex
\documentclass[sigconf,natbib=false]{acmart}

\usepackage{dblfnote}
\AtBeginDocument{%
  }

\acmYear{2024}\copyrightyear{2024}
\acmConference[HPDC '24]{The 33rd International Symposium on High-Performance Parallel and Distributed Computing}{June 3--7, 2024}{Pisa, Italy}
\acmBooktitle{The 33rd International Symposium on High-Performance Parallel and Distributed Computing (HPDC '24), June 3--7, 2024, Pisa, Italy}
\acmDOI{10.1145/3625549.3659133}
\acmISBN{979-8-4007-0413-0/24/06}

\setcopyright{none}
\renewcommand\footnotetextcopyrightpermission[1]{} 
\RequirePackage[
  datamodel=acmdatamodel,
  style=acmnumeric,
  ]{biblatex}

\usepackage{soul,color}

\usepackage[acronym,toc,shortcuts]{glossaries}

\newacronym{5G}{5G}{Fifth Generation}
\newacronym{6G}{6G}{Sixth Generation}
\newacronym{3GPP}{3GPP}{The 3rd Generation Partnership Project}
\newacronym{AAA}{AAA}{Authentication, Authorization and Accounting}
\newacronym{ACT}{ACT}{Actuator}
\newacronym{AD-SAL}{AD-SAL}{API-Driven Service Abstraction Layer}
\newacronym{AE}{AE}{Analytical Engine}
\newacronym{AI}{AI}{Artificial Intelligence}
\newacronym{AP}{AP}{Access Point}
\newacronym{API}{API}{Application Programming Interface}
\newacronym{CCL}{CCL}{fully centralized SLA-constrained deep learning}
\newacronym{CL}{CL}{Closed Loop}
\newacronym{CNF}{CNF}{Cloud Native Function}
\newacronym{COMS}{COMS}{Common Online Memory Store}
\newacronym{CQI}{CQI}{Channel Quality Indicator}
\newacronym{CPU}{CPU}{Central Processing Unit}
\newacronym{CU}{CU}{Centralized Unit}
\newacronym{DE}{DE}{Decision Engine}
\newacronym{DRL}{DRL}{Decentralized Deep Reinforcement Learning}
\newacronym{DU}{DU}{Distributed Unit}
\newacronym{eMBB}{eMBB}{enhanced Mobile Broadband}
\newacronym{EEM}{EEM}{Embedded Element Manager}
\newacronym{ENI}{ENI}{Experiential Networked Intelligence}
\newacronym{ETSI}{ETSI}{European Telecommunications Standards Institute}
\newacronym{FL}{FL}{Federated Learning}
\newacronym{IoT}{IoT}{Internet of Things}
\newacronym{KEM}{KEM}{Key Encapsulation Mechanism}
\newacronym{KMS}{KMS}{Key Management System}
\newacronym{KPI}{KPI}{Key Performance Indicator}
\newacronym{LCM}{LCM}{Life Cycle Management}
\newacronym{MANO}{MANO}{Management and Orchestration}
\newacronym{MEC}{MEC}{Multi-Access Edge Computing}
\newacronym{MIMO}{MIMO}{Multiple-Input Multiple-Output}
\newacronym{ML}{ML}{Machine Learning}
\newacronym{MS}{MS}{Monitoring System}
\newacronym{NFV}{NFV}{Network Function Virtualization}
\newacronym{NIST}{NIST}{National Institute of Standards and Technology }
\newacronym{NMSE}{NMSE}{Normalized Mean Squared Error}
\newacronym{NS}{NS}{Network Slice}
\newacronym{OSS}{OSS}{Operational Subsystem}
\newacronym{QoS}{QoS}{Quality of Service}
\newacronym{QKD}{QKD}{Quantum Key Distribution}
\newacronym{PQC}{PQC}{Post-Quantum Cryptography}
\newacronym{PRB}{PRB}{Physical Resource Block}
\newacronym{RAN}{RAN}{Radio Access Network}
\newacronym{REST}{REST}{Representational State Transfer}
\newacronym{SDN}{SDN}{Software Defined Networks}
\newacronym{SDO}{SDO}{Standards Development Organization}
\newacronym{SF}{SF}{Sampling Function}
\newacronym{SLA}{SLA}{Service Level Agreement}
\newacronym{StFL}{StFL}{Statistical Federated Learning}
\newacronym{TFCO}{TFCO}{TensorFlow Constrained Optimization}
\newacronym{TRP}{TRP}{Transmission/Reception Point}
\newacronym{TSDB}{TSDB}{Time Series Database}
\newacronym{UE}{UE}{User Equipment}
\newacronym{UPF}{UPF}{User Plane Function}
\newacronym{VNF}{VNF}{Virtual Network Function}
\newacronym{ZSM}{ZSM}{Zero-touch network Service Management}

\begin{document}

\title{Exploring  Post Quantum Cryptography with Quantum Key Distribution for Sustainable Mobile Network Architecture Design}


\author{Sanzida Hoque}
\affiliation{%
  \institution{Florida Institute of Technology}
  \streetaddress{1 Th{\o}rv{\"a}ld Circle}
  \city{Melbourne, FL}
  \country{USA}
  \postcode{32901}}
\email{Email: shoque2023@my.fit.edu}

\author{Abdullah Aydeger}
\affiliation{%
  \institution{Florida Institute of Technology}
  \streetaddress{1 Th{\o}rv{\"a}ld Circle}
  \city{Melbourne, FL}
  \country{USA}
  \postcode{32901}}
\email{Email: aaydeger@fit.edu}

\author{Engin Zeydan}
\affiliation{%
  \institution{Centre Tecnològic de Telecomunicacions de Catalunya (CTTC/CERCA), Castelldefels, Spain}
  \city{}
  \country{}
  \postcode{08860}}
  \email{Email: ezeydan@cttc.es}

\begin{abstract}

The proliferation of mobile networks and their increasing importance to modern life, combined with the emerging threat of quantum computing, present new challenges and opportunities for cybersecurity. This paper addresses the complexity of protecting these critical infrastructures against future quantum attacks while considering operational sustainability. We begin with an overview of the current landscape, identify the main vulnerabilities in mobile networks, and evaluate existing security solutions with new post-quantum cryptography (PQC) methods. We then present a quantum-secure architecture with PQC and Quantum Key Distribution (QKD) tailored explicitly for sustainable mobile networks and illustrate its applicability with several use cases that emphasize the need for advanced protection measures in this new era. In addition, a comprehensive analysis of PQC algorithm families is presented, focusing on their suitability for integration in mobile environments, with particular attention to the trade-offs between energy consumption and security improvements. Finally, recommendations for strengthening mobile networks against quantum threats are provided through a detailed examination of current challenges and opportunities.

\end{abstract}


\maketitle


\section{Introduction}


In recent years, the rapid development of quantum computing in the field of cyber security has caused both excitement and concern. While quantum computing promises unprecedented computing power and the ability to solve complex problems previously thought unfeasible, it also poses significant challenges to the security infrastructure that underpins our digital world. As quantum technologies continue to mature, understanding the security implications of quantum computing is of paramount importance. In response to this looming threat, researchers and industry experts have turned to innovative solutions such as \ac{PQC} and \ac{QKD} to protect the confidentiality and integrity of digital communications in the quantum age.

In the context of developing a sustainable architecture for mobile networks, \ac{PQC} and \ac{QKD} offer crucial solutions to achieve long-term security without jeopardizing the environmental and operational sustainability of these networks. The crucial role of QKD in a sustainable architecture for mobile networks lies in its ability to ensure secure communication through the principles of quantum physics, which theoretically guarantees security indefinitely without the need for frequent updates or upgrades that consume additional resources. PQC, on the other hand, offers a directly applicable approach to securing mobile networks against quantum threats. It is a new field of research that deals with the development of cryptographic algorithms that are resistant to attacks from quantum computers. Therefore, the combination of QKD's unconditionally secure communication channels with PQC's quantum-resistant encryption and key exchange mechanisms enables the development of secure mobile networks that are resistant to both current and future threats without the need for frequent, resource-intensive upgrades. By integrating QKD and PQC, networks can leverage the strengths of both technologies — the unmatched security of QKD for key distribution and the flexibility and efficiency of PQC for widespread cryptographic applications. This balanced approach can enable the deployment of secure, sustainable mobile networks that can adapt to changing security requirements while minimizing their environmental footprint.

\footnote{This paper has been accepted for publication by HPDC '24. The copyright is with the ACM and the final version will be published by the ACM}

In this paper, we leverage PQC and QKD to design an integrated novel mobile network software defined architecture that enhances security against potential quantum attacks and ensures that the security solutions can be adapted to future advances in quantum computing. By exploring different PQC algorithms and the potential of QKD to reduce the need for frequent infrastructure upgrades, we offer a new perspective on building future-proof and sustainable mobile networks. At the end of the paper, we also address the current limitations and challenges in an integrated PQC/QKD network environment and highlight some opportunities and recommendations.


\input{related}

\input{background}

\input{usecase}

\input{pqcmobile}

\section{Challenges, Opportunities and Recommendations}



\input{limitations}

\section{Conclusions and Future Directions}


The evolution of mobile technologies, in particular the transition from 5G to 6G, will have a profound impact on the application of \ac{PQC} and \ac{QKD}. As we move into this new quantum era, mobile network operators need to innovate and collaborate to overcome the challenges and harness the power of PQC and QKD in their infrastructure. In this work, the integration of PQC and QKD in the architecture of mobile networks has been studied, security vulnerabilities have been investigated, PQC algorithm families have been compared and a quantum-secure architecture tailored to the mobile context has been proposed. Looking ahead, several key areas such as hybrid cryptographic systems, quantum threat modeling, risk assessment, and cross-domain applications are considered crucial for further research to ensure the seamless integration of PQC and QKD into mobile infrastructures and address the evolving cybersecurity threat landscape.

\printbibliography

\end{document}

%% file: related.tex
\section{Related Works}
The traditional Public Key Infrastructure (PKI) is based on cryptographic algorithms such as Rivest–Shamir–Adleman (RSA) and Elliptic curve cryptography (ECC), which are vulnerable to attacks by quantum computers. PQC and QKD solutions are the response and preparation for the Q-Day~\cite{ford2023quantum} when powerful quantum computers may break these known algorithms using Shor's algorithm~\cite{kumar2020post}. In this section, the recent research focuses on developing PQC and QKD, especially for mobile networks, are discussed. Some of these works focus solely on PQC, QKD, or both as complementing components for establishing quantum-secure communication. This study~\cite{zeydan2022recent} investigates the recent advancements in PQC algorithms in networking and telecommunications and provides a comprehensive list of issues and solutions for securing communication networks in the post-quantum era. In another study~\cite{partala2021post}, the authors explore secure post-quantum options for key establishment, public-key encryption, and digital signatures, as well as discuss their features and impact on future 6G network performance. The authors in ~\cite{abdallah2024physical} introduce a new encryption scheme that uses lattice cryptography, a promising area of research in PQC, for the physical layer of 6G networks. This scheme aims to address the security challenges that arise due to the extensive data traffic in these networks. NIST's post-quantum algorithm CRYSTALS-Kyber and Round 4 candidate \glspl{KEM} are used to evaluate the proposed protocol. The paper in~\cite{liu2021post} proposes a new double authentication preventing ring signature-based privacy protection scheme (PRSG) using a post-quantum secure approach to improve security and privacy in the 6G cybertwin system. The approach involves creating an accumulator using a chameleon hash function and a general double authentication preventing ring signature (DAPRS) with an efficient zero-knowledge membership proof. KEMSUCI~\cite{ulitzsch2022post}, a new post-quantum secure technique for Subscription Concealed Identifier (SUCI) calculation, is proposed by Ulitzsch et al. The proposed technique can utilize any \glspl{KEM} from the NIST call for standardizing PQC. Among the candidates, Kyber and Saber are assessed to be the best fit to address quantum attacks. This approach can work with 5G and 6G networks and has less communication overhead than other Subscription Permanent Identifier (SUPI) protection methods.





Recently, there has been a growing interest in integrating quantum-based approaches into mobile networks \cite{hanna2023performance}. A hybrid approach for quantum-safe communication is proposed by SKT\footnote{Online: https://www.sktelecom.com/en/press/press\_detail.do?idx=1579, Available: March 2024.}, which includes QKD for the backbone of the mobile network and PQC for the UE side of the infrastructure. The paper~\cite{brauer2023linking} outlines effective methods for quantum-safe key exchange between three major European QKD testbeds, which differ in network architecture, functionalities and management, reflecting different ways to operate the infrastructure for a telecommunication company. A pan-European E2E key exchange is being set up using SDN for QKD. Another study~\cite{martin2024service} presents a test environment that allows researchers to emulate attacks on the QKD network without the need for physical quantum equipment. The solution complies with ETSI standards, uses open source technologies and is validated through trials at the 5G Telef\'{o}nica Open Network Innovation Center. In another way, the paper \cite{geitz2023hybrid} introduces hybrid protocols using QKD and PQC in the OpenQKD testbed to improve the security of QKD nodes. The authors have developed a quantum secure key management system that uses PQC algorithms. Applications can request arbitrary encryption keys from the KMS and combine them arbitrarily to obtain more secure cryptographic keys. The paper presents hybrid protocols that use QKD and PQC in the OpenQKD testbed to enhance the security of QKD nodes. The authors have developed a quantum secure key management system that uses PQC algorithms. . 

As we have summarized here, the previous work in PQC/QKD has dealt with various aspects of communication systems. However, not much has been discussed about the integration of these systems into mobile networks. Also, the sustainability comparisons of possible architectures has not yet been investigated.  In comparison to previous works, we propose a novel architecture for mobile networks and make some suggestions for the best algorithmic approaches that can be used within this context.


%% file: background.tex
\section{Background on Mobile Network Security and PQC/QKD}

In this section, we provide the vulnerabilities in the mobile networks and discuss the potential solutions utilizing PQC and QKD.




\subsection{Security Vulnerabilities }

\textbf{Authentication and Data Protection:} Mobile networks are vulnerable to man-in-the-middle attacks, where an attacker intercepts and modifies the communication between two parties. PQC algorithms can strengthen authentication mechanisms by providing robust digital signatures or key exchange protocols that are resistant to quantum attacks, mitigating the risk of unauthorized access to network resources. PQC algorithms can also provide strong encryption techniques to protect the confidentiality of data, ensuring that sensitive information remains secure even if it is intercepted by an attacker with access to quantum computers.

\textbf{Signaling Security:} Signaling protocols in mobile networks, such as those used for connection establishment, authentication, and session management, are vulnerable to spoofing and tampering attacks. PQC-based digital signatures and authentication mechanisms can help to verify the integrity and authenticity of signaling messages and prevent unauthorized changes or impersonation attempts by malicious entities. 

\textbf{Key Management:} Mobile networks rely on secure key management to generate and manage cryptographic keys for encryption, authentication, and secure communication. PQC algorithms can improve the security of key management by providing quantum-resistant key exchange protocols and key generation mechanisms, thus reducing the risk of key compromise by quantum-assisted attacks.

\textbf{Network Infrastructure Security:} Mobile network infrastructure, including base stations, routers, and core network elements, is vulnerable to Denial of Service (DoS) attacks that disrupt or impair network services. PQC algorithms can help mitigate the impact of DoS attacks by strengthening authentication and access control mechanisms and limiting the ability of attackers to exploit vulnerabilities in network protocols and infrastructure.

\textbf{Privacy Preservation:} Mobile networks collect and transmit location data, which raises privacy and surveillance concerns. PQC-based encryption techniques can protect location data from unauthorized access, maintain user privacy and prevent tracking and profiling by attackers.


\subsection{Traditional and  PQC-based Security Solutions}
PQC can close various security gaps in mobile networks by providing robust authentication, data protection, signal security, key management, infrastructure security, privacy protection, and future-proof mechanisms against quantum computing threats. By integrating PQC into the security architectures of mobile networks, organizations can improve the overall resilience and trustworthiness of mobile communications and services.
Next, we discuss some potential PQC algorithm candidates for the wide adaptation to the current systems.

\subsubsection{PQC algorithms}
PQC candidates contain several algorithmic families. Each of these families is based on mathematical challenges that are difficult to solve with classical and quantum computing systems. The goal of PQC is to protect cryptographic primitives such as encryption, digital signatures, and key exchange against potential threats from quantum computers. These families offer different approaches to achieving this goal. Fig~\ref{fig:tree} shows the seven finalists and eight alternates announced by NIST in the third round of candidates, grouped by each family. The algorithms outlined in bold are the four candidates for standardization, and the boxes with a hashed pattern are the candidates for the \glspl{KEM} algorithms that will advance to the round 4. Here's a brief overview of the major PQC families~\cite{kumar2020post, wang2021hardware, dam2023survey}:

\vspace{-0.1in}
\begin{figure}[htp!]
\centering
\includegraphics[width=\linewidth]{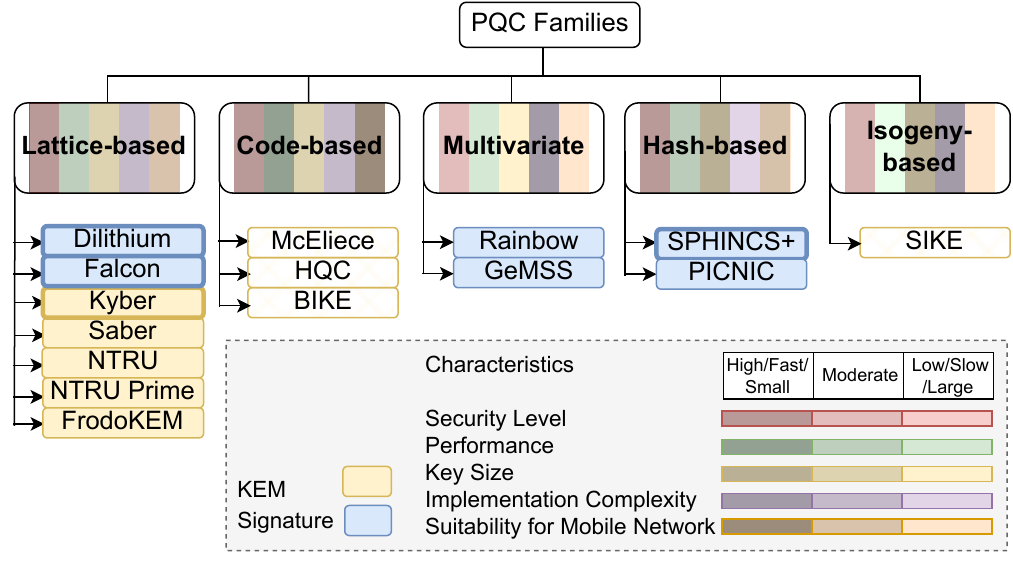}
\vspace{-0.3in}
\caption{PQC Algorithms Categorization.} 
\label{fig:tree}
\vspace{-0.1in}
\end{figure}

\textbf{Lattice-Based Cryptography:} The basis of lattice-based cryptography lies in the difficulty of solving problems with lattice structures in spaces with a high number of dimensions \cite{eich2023quantum}. This family is known for its efficiency in the development of encryption and digital signature algorithms \cite{nejatollahi2019post}. Learning With Errors (LWE) and its variations include schemes such as Kyber for key exchanges, and Falcon and Dilithium for digital signatures, which have been selected by NIST for standardization. 

\textbf{Code-Based Cryptography:}This family is based on the difficulty of decoding randomly generated linear codes~\cite{balamurugan2021post}. The Classic McEliece, which was also selected by NIST (including HQC and BIKE as candidates for the fourth round), is the best known algorithm in this category and provides a secure communication architecture against quantum attacks. Code-based cryptography is known for its high security and effectiveness, but usually requires comparatively larger key sizes.

\textbf{Multivariate Cryptography: } This family is based on the difficulty of solving systems of multivariate quadratic equations over a finite field~\cite{ikematsu2023recent}. NIST considered the use of multivariate cryptography to create secure digital signatures. However, no multivariate scheme was selected for standardization in the last round of evaluation, in part due to concerns about security and performance~\cite{alagic2022status}. 

\textbf{Hash-Based Cryptography:} The family of PQC-dependent hash functions~\cite{dam2023survey} and known for generating secure digital signatures. Hash-based signatures are a type of post-quantum encryption that is not vulnerable to quantum attacks. SPHINCS+ is a hash-based approach to digital signatures selected by NIST. The system is valued for its high level of security, which does not depend on unproven mathematical challenges and does not require a large key size.

\textbf{Isogeny-Based Cryptography:} This family uses the properties of elliptic curves and their isogenies to develop cryptographic systems. SIKE (Supersingular Isogeny Key Encapsulation)\cite{elkhatib2020efficient} was selected for further evaluation in the fourth round and shows the continued investigation and promise of isogeny-based methods to protect future communications, but it is already broken~\cite{Chen}.

\subsubsection{Comparison with Traditional Approaches}

Existing cryptographic solutions such as RSA and ECC can be compared with potential \ac{PQC}-based solutions such as Dilithium and Kyber in terms of several key features. 
In terms of security levels, RSA and ECC, although widely used, are vulnerable to quantum attacks because they are based on mathematical problems that quantum computers can solve efficiently. Dilithium and Kyber, on the other hand, were specifically designed to resist such attacks and provide robust security in anticipation of quantum computing advances.
However, there are also performance trade-offs to consider. Regarding computational speed, RSA and ECC generally outperform PQC-based solutions such as Dilithium and Kyber. The latter may require more computing resources and energy, leading to slower performance, especially in resource-constrained environments. The key size is another distinguishing feature. RSA and ECC often have smaller key sizes, which can be advantageous for efficiency and storage. In contrast, PQC-based solutions may require larger key sizes, which affects storage requirements and transmission overhead. 
RSA and ECC benefit from well-established interoperability with existing systems and protocols, contributing to their broad acceptance. However, PQC-based solutions such as Dilithium and Kyber may require further development and standardization efforts to achieve comparable interoperability with existing cryptographic infrastructures. Therefore, stakeholders must carefully consider the impact on performance, key sizes, and interoperability requirements when evaluating their suitability for specific applications.

%% file: usecase.tex
\begin{figure}[htp!]
\centering
\includegraphics[width=\linewidth]{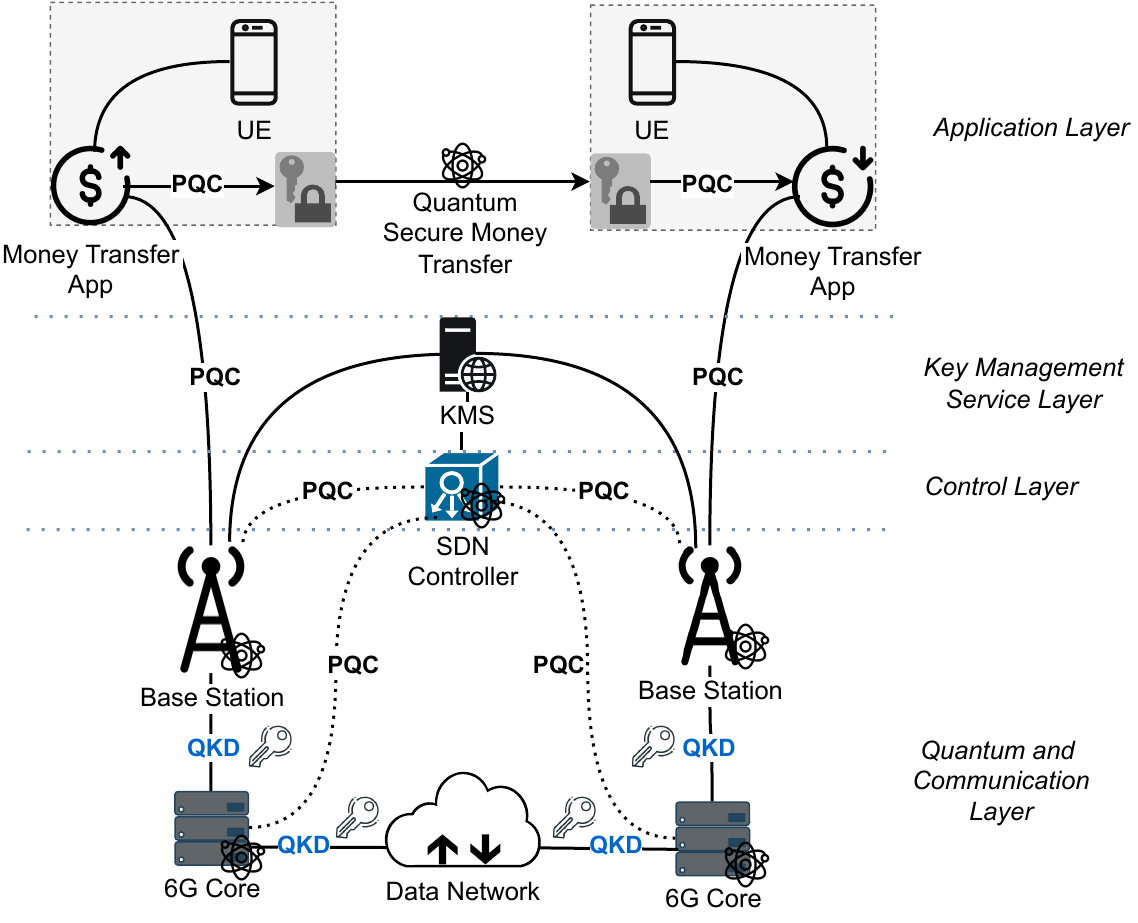}
\vspace{-0.25in}
\caption{Quantum Secure Communication between peers when utilizing mobile network.} 
\label{fig:pqc_2}
\vspace{-0.15in}
\end{figure}

\section{Quantum Secure Architecture and Use Cases for Mobile Networks}
\label{sec:arch}
In this section, we describe our proposed architecture and present some possible use cases for mobile networks. We also explain the updated packet exchange required to support such a quantum-safe mobile network architecture.

\subsection{Proposed Architecture}

Our proposed diagram is displayed in Fig. \ref{fig:pqc_2}, where the quantum-safe communication between peers is presented using a mobile network. The components of this architecture are explained as follows:  

\textbf{(i) User Equipment (UE):} This is the typical device that the users can connect to the 6G networks with. UEs are enabled to support PQC algorithms for the key exchange purposes. We consider a phone application (i.e., a Money Transfer App) for the data traffic via the mobile network.  

\textbf{(ii) Base Station (BS) and Beyond 5G/6G Core (B5G/6GC):} These are the traditional components of the mobile network backbone that are responsible for handling high-level network operations such as user authentication, session management, mobility management and interfaces to external networks. We are considering the installation of quantum gateways for each of these devices. The Quantum Gateway is an advanced network device that utilises the principles of quantum computing to enhance security and communication capabilities. It serves as an interface for the quantum communication channels required for QKD. Thus, BS and B5G/6GC can enable various quantum secure communications, such as (1) PQC-based key exchange via UE, (2) QKD between BS and B5G/6GC, (3) QKD for the B5G/6GC to Data Network.   

\textbf{(iii) SDN Controller/KMS:} Network traffic is managed via the SDN Controller and the keys are forwarded to the KMS by it.  

\textbf{(iv) Data network:} This is where the data leaves the mobile operator's network and goes to the Internet. We propose the QKD is supporte (i.e., with Quantum Gateways) until the last routers of the mobile operator's network.

\begin{figure}[htp!]
\centering
\vspace{-0.1in}
\includegraphics[width=.8\linewidth]{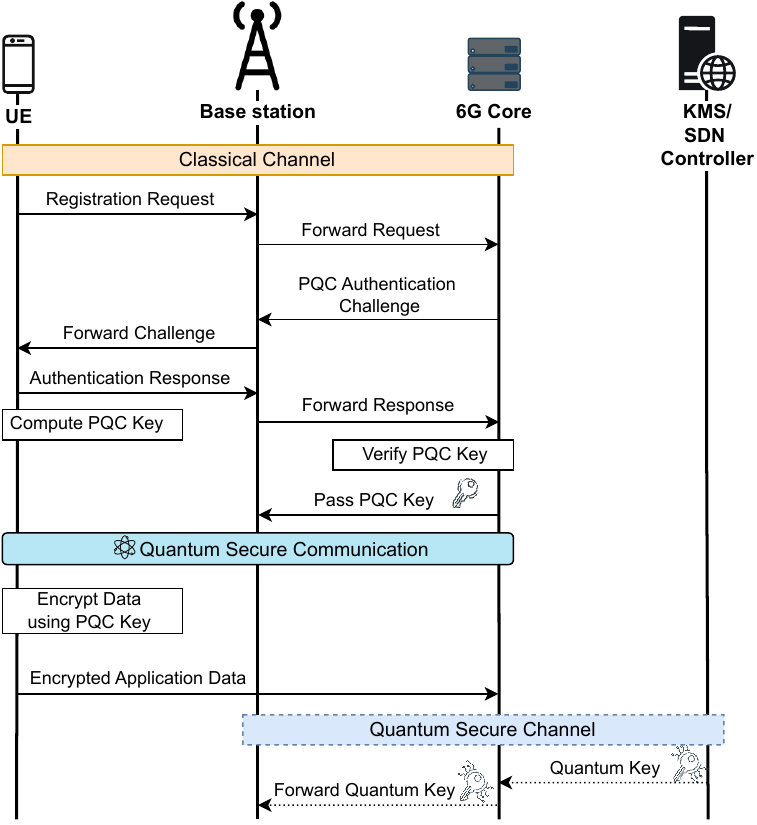}
\caption{PQC/QKD Packet Exchanges between UE to 6G Core} 
\vspace{-0.2in}
\label{fig:flowchart}
\end{figure}

\subsection{Flowchart for Encrypted Data Transmission}

Fig. \ref{fig:flowchart} shows the flowdiagram for a PQC-QKD integrated mobile infrastructure. In this diagram, the \ac{KMS}/ \ac{SDN} controller manages and stores the keys. The packet exchange for UE authentication is the improved version of EPC-AKA authentication and key exchange \cite{etsi} to support PQC/QKD in the mobile network. When the UE wants to join the network for the first time, it sends a registration request (i.e. an AUTH request), which is forwarded to the 6G core. Using the UE's identification information in this request, the 6G Core generates a PQC Authentication Challenge, including RAND, and sends it back to the UE. After receiving this challenge, the UE responds with the Authentication Response and simultaneously generates the PQC Session Key (i.e. a symmetric key), which is later used for data traffic. After receiving the challenge response, 6G Core verifies if the message is as expected and then generates the PQC session key to be used for the traffic of that particular UE. The PQC session key is also forwarded to the base station to which the UE connects. After the session keys have been generated via PQC, the application data is encrypted with this key. In the backbone of the mobile network, this traffic is routed through the Quantum Secure Channel, which relies on QKD exchange for the reliability, integrity and security of these messages.

\subsection{Potential Use Cases in Mobile Networks}

In mobile networks, PQC and QKD algorithms can be embedded in several places: (1) between UEs and BS traffic, (2) between BS and 6GC, and (3) between 6GC and data networks (where the data leaves the mobile operator), as shown in Fig. \ref{fig:pqc_2}. In this paper, we consider and compare each of these three approaches and recommend the best scenarios and PQC/QKD integration ideas. 

For data traffic between UEs and BSs, PQC (e.g. CRYSTALS-Kyber) can be used as a public/private key algorithm by having the UEs (i.e., mobile phones) perform these key generation operations every time they connect to a new BS. Given the mobility of UEs, these operations are likely to be repeated very often. However, a UE can use a key obtained via CRYSTALS-Kyber key exchange for all its traffic. The main idea behind this is that we encrypt the traffic only between UE and BS and not end-to-end for the proposed PQ-based security in the mobile networks. End-to-end encryption could still run on such infrastructure and provide an additional layer of security. However, this is outside the scope of this paper. Therefore, it is not necessary for the UEs to have different keys unless the effective BS has been changed. 

In the second option, where QKD is used for the traffic between BS and 6GC, we do not provide direct protection for the UEs' traffic against possible quantum computing threats. However, we facilitate encryption for the backend of the mobile network, where the threat model includes different attack vectors (e.g., insider threats, virtualization-based threats, etc.). In this approach, one-time key generation could be used for a large amount of traffic, as many UEs could be connected to this BS, and all their traffic could go through this BS. Also, since the BS and the 6GC are static, the keys for the encryption sessions do not need to be updated frequently. Therefore, this approach is much lighter and more stable than the first one. However, this approach requires additional hardware, the Quantum Gateway, as defined in Section \ref{sec:arch}, which supports quantum communication between BS and 6GC.  As a third possibility, we can facilitate the QKD between 6GC and DN. This method is similar to the second technique in terms of the frequency of key generation and the number of devices involved. However, this approach is directly effective against potential threats in a much larger environment (i.e., the entire Internet). The importance of this approach surpasses that of the first two, considering the possibility of attacks via quantum computers.  Some potential use cases where PQC/QKD can be considered and deployed in mobile networks are: 

\textbf{(i) Secure communication protocols:} PQC algorithms can be integrated into mobile communication protocols (i.e. 6G) to provide secure encryption and authentication mechanisms. This ensures that sensitive information exchanged between mobile devices and the network infrastructure remains confidential and protected from quantum attacks. 

\textbf{(ii) Authentication of user devices:} When the UE joins the network for the first time via the BS, packets must be exchanged to create the PQC keys, as shown in Fig. \ref{fig:flowchart}. This process still relies on the master key (i.e. the unique identifier of the UE), which needs to be defined and embedded in the UEs. 

\textbf{(iii) Encryption of user equipment data:} After initialization of the UE with the mobile network via the BS, the UE and the BS generate their keys via PQC (e.g. CRYSTALS-Kyber). The data traffic, e.g. a money transfer app as shown in Fig. \ref{fig:pqc_2}, is encrypted with the PQ keys and offers security against possible quantum attacks.

Overall, the use of PQC and QKD in mobile networks offers the potential to increase the security and resilience of mobile communications and services in the face of emerging quantum computing threats. As quantum computing technology advances, the integration of PQC and QKD will become increasingly important to ensure the confidentiality, integrity and authenticity of mobile network operations and transactions.	



%% file: pqcmobile.tex

\section{PQC algorithms for mobile networks}
In this section, we first compare and contrast different PQC algorithm families. Next, we discuss the energy consumption for each of them for the sustainable mobile network environments. Finally, we explore and suggest specific PQC protocols for a few mobile network use cases.

\subsection{Comparisons of PQC Algorithm Families}

Table \ref{PQC_comp} compares different families of PQC algorithms based on various characteristics following ~\cite{alagic2022status, Chen, dam2023survey, hession20191, Moody}. \textbf{In terms of security level}, both lattice-based and code-based algorithms provide a high level of security, i.e., they are resistant to attacks from both classical and quantum computers (see \cite{bernstein2017post} for security foundations of these families). In addition to having solid theoretical foundations and extensive security analyses, hash-based algorithms are also resistant to quantum attacks. This is crucial for mobile networks where the transmission of sensitive data requires robust encryption to protect against potential threats. \textbf{In terms of performance,} code-based algorithms are characterized by their fast performance, which makes them attractive for mobile networks where efficiency is important. Lattice-based and hash-based algorithms offer moderate performance, while multivariate and isogeny-based algorithms tend to be slower. In mobile networks, performance is often prioritized to ensure smooth and responsive communication for users. Therefore, code-based algorithms are particularly attractive in this context.

\begin{table*}[htp!]
\centering
\caption{Comparison of PQC Algorithm Families}
\vspace{-0.1in}
\label{tab:crypto_comparison}
\begin{tabular}{|l|c|c|c|c|c|}
\hline
\multicolumn{1}{|c|}{\begin{tabular}[c]{@{}c@{}} \textbf{Feature}  \end{tabular}}  & 
\multicolumn{1}{|c|}{\begin{tabular}[c]{@{}c@{}} \textbf{Hash} \textbf{based} \end{tabular}}  & 
\multicolumn{1}{|c|}{\begin{tabular}[c]{@{}c@{}} \textbf{Isogeny}  \textbf{based} \end{tabular}}  & 
\multicolumn{1}{|c|}{\begin{tabular}[c]{@{}c@{}} \textbf{Lattice}  \textbf{based} \end{tabular}}  & 
\multicolumn{1}{|c|}{\begin{tabular}[c]{@{}c@{}} \textbf{Code}  \textbf{based} \end{tabular}} & 
\multicolumn{1}{|c|}{\begin{tabular}[c]{@{}c@{}} \textbf{Multivariate}  \textbf{} \end{tabular}} \\ \hline 
\multicolumn{1}{|c|}{\begin{tabular}[c]{@{}c@{}} Security level \end{tabular}} & High & Moderate & High & High & Moderate \\ \hline
\multicolumn{1}{|c|}{\begin{tabular}[c]{@{}c@{}} Performance \end{tabular}}  & Moderate & Slow & Moderate & Fast & Slow \\ \hline
\multicolumn{1}{|c|}{\begin{tabular}[c]{@{}c@{}} Key size \end{tabular}} 
 & Small & Small & Moderate & Moderate & Large \\ \hline
\multicolumn{1}{|c|}{\begin{tabular}[c]{@{}c@{}} Implementation complexity \end{tabular}} 
 & Moderate & High & Moderate & Moderate & High \\ \hline
\multicolumn{1}{|c|}{\begin{tabular}[c]{@{}c@{}} Suitability  for mobile networks \end{tabular}}  & Moderate & Low  & Moderate & High & Low \\ \hline
\end{tabular}
\label{PQC_comp}
\vspace{-0.1in}
\end{table*}

\textbf{In terms of key size}, code-based algorithms usually have a moderate key size, which strikes a balance between security and efficiency (although some schemes like Classic McEliece have notably large key sizes for high security). Lattice-based algorithms generally have a moderate key size as well. Multivariate algorithms require larger keys, which can be a challenge for mobile networks in terms of storage and processing requirements. Hash-based and isogeny-based algorithms have smaller key sizes but may compromise performance to achieve this reduction. \textbf{In terms of implementation complexity}, both lattice-based, code-based, and hash-based algorithms are moderately complex, so they can be used in mobile networks with adequate resources and expertise. However, isogeny-based and multivariate algorithms have comparatively higher implementation complexity, which could make integration into mobile network systems challenging. \textbf{In terms of suitability for mobile networks}, code-based algorithms prove to be particularly suitable for mobile networks due to their combination of high security, fast performance, moderate key size, and manageable implementation complexity. Lattice-based algorithms also provide a reasonable balance of these factors, although they may be less optimal in terms of performance than code-based algorithms. Hash-based algorithms with small key sizes can be a moderately suitable option. Isogeny-based and multivariate algorithms, with their slower performance and higher implementation complexity, are less suitable for mobile networks where efficiency and resource constraints play an important role.

To summarise, code-based and lattice-based PQC algorithms seem to be more suitable for use in mobile networks than multivariate algorithms. However, the specific choice between these algorithms depends on network requirements, priorities, and available resources.

\subsection{Energy and Security Trade-offs}

PQC implementations in mobile environments will aim to strike a balance between energy efficiency and security by using optimized algorithms and techniques to provide robust cryptographic protection while minimizing the impact on the mobile device's resources \cite{lakhan2023comparative}. As PQC continues to evolve and standardize, we can expect to see further advances in energy-efficient implementations and improved security techniques tailored specifically for mobile networks. One approach to improving the energy efficiency of PQC implementations for mobile environments is to carefully select algorithms that provide a good balance between security and computational overhead. Algorithms such as NTRUEncrypt, a lattice-based method known for its relatively low computational overhead compared to some other PQC methods, can be favored for mobile applications where energy efficiency is critical. Another strategy is to optimize the PQC algorithms for mobile platforms to minimize energy consumption. This may include techniques such as optimizing algorithms, using hardware acceleration (if available), and minimizing memory usage to reduce power consumption during cryptographic operations.

Another approach to improving security is for organizations such as the NIST (National Institute of Standards and Technology) to examine various PQC algorithms for their security properties as part of standardization efforts. PQC implementations in mobile environments can benefit from the adoption of standardized algorithms that have undergone rigorous testing and analysis to ensure their security against classical and quantum attacks. Mobile devices are also vulnerable to side-channel attacks, where an attacker exploits information leaked during cryptographic operations (e.g. power consumption, timing). Implementations of PQC algorithms for mobile environments must include countermeasures to mitigate such attacks, e.g. constant-time algorithms, randomization and secure key storage mechanisms. Apart from the choice of cryptographic algorithms, ensuring overall security in mobile environments is linked to the implementation of secure protocols and best practices. This includes key management mechanisms, secure communication channels and protection against various threats such as man-in-the-middle attacks and malware targeting mobile devices.

\subsection{Application Specific PQC Algorithm Comparisons}


\begin{table*}[htp!]
\centering
\caption{Comparison of cryptographic schemes for different domains in mobile networks.}
\label{tab:cryptoschemes_comparison}
\vspace{-0.1in}
\begin{tabular}{|l|l|l|l|l|}
\hline
\multicolumn{1}{|c|}{\begin{tabular}[c]{@{}c@{}} \textbf{Domains} \end{tabular}}   & \multicolumn{1}{|c|}{\begin{tabular}[c]{@{}c@{}} \textbf{Scheme 1} \end{tabular}}  & \multicolumn{1}{|c|}{\begin{tabular}[c]{@{}c@{}} \textbf{Scheme 2} \end{tabular}} & \multicolumn{1}{|c|}{\begin{tabular}[c]{@{}c@{}} \textbf{Scheme 3} \end{tabular}} & \multicolumn{1}{|c|}{\begin{tabular}[c]{@{}c@{}} \textbf{Scheme 4} \end{tabular}} \\ \hline
\multicolumn{1}{|c|}{\begin{tabular}[c]{@{}c@{}} User authentication and signaling security \end{tabular}}  & \multicolumn{1}{|c|}{\begin{tabular}[c]{@{}c@{}} Crystals-DILITHIUM \end{tabular}}   & \multicolumn{1}{|c|}{\begin{tabular}[c]{@{}c@{}} Falcon \end{tabular}}   & \multicolumn{1}{|c|}{\begin{tabular}[c]{@{}c@{}} Rainbow \end{tabular}}   &  \multicolumn{1}{|c|}{\begin{tabular}[c]{@{}c@{}} SIKE \end{tabular}}  \\ \hline
\multicolumn{1}{|c|}{\begin{tabular}[c]{@{}c@{}} Data encryption and privacy preservation \end{tabular}}   & \multicolumn{1}{|c|}{\begin{tabular}[c]{@{}c@{}} Kyber \end{tabular}}  & \multicolumn{1}{|c|}{\begin{tabular}[c]{@{}c@{}} NTRU \end{tabular}} & \multicolumn{1}{|c|}{\begin{tabular}[c]{@{}c@{}} BIKE \end{tabular}} &   \multicolumn{1}{|c|}{\begin{tabular}[c]{@{}c@{}} HQC \end{tabular}}  \\ \hline
\multicolumn{1}{|c|}{\begin{tabular}[c]{@{}c@{}} Network management  and infrastructure security \end{tabular}} & \multicolumn{1}{|c|}{\begin{tabular}[c]{@{}c@{}} Kyber \end{tabular}}  & \multicolumn{1}{|c|}{\begin{tabular}[c]{@{}c@{}} NTRU \end{tabular}}  & \multicolumn{1}{|c|}{\begin{tabular}[c]{@{}c@{}} Classic McEliece  \end{tabular}}  & \multicolumn{1}{|c|}{\begin{tabular}[c]{@{}c@{}} SIKE \end{tabular}}  \\ \hline
\multicolumn{1}{|c|}{\begin{tabular}[c]{@{}c@{}} Key management \end{tabular}}
& \multicolumn{1}{|c|}{\begin{tabular}[c]{@{}c@{}} Crystals-KYBER \end{tabular}} & \multicolumn{1}{|c|}{\begin{tabular}[c]{@{}c@{}} NTRU \end{tabular}} & \multicolumn{1}{|c|}{\begin{tabular}[c]{@{}c@{}} SABER  \end{tabular}}  & \multicolumn{1}{|c|}{\begin{tabular}[c]{@{}c@{}} BIKE \end{tabular}}  \\ \hline
\end{tabular}
\vspace{-0.1in}
\end{table*}

Table \ref{tab:cryptoschemes_comparison} shows the comparison of PQC schemes for different areas in mobile networks. 

\textbf{In terms of user authentication and signaling security,} digital signature algorithms such as Crystals-DILITHIUM, Falcon, and Rainbow are crucial. Digital signatures provide integrity, non-repudiation, and authenticity, which are critical for authenticating users and securing signaling in mobile networks. Crystals-DILITHIUM is known for its balance between signature size and security, making it efficient for authentication. Falcon offers smaller signatures and fast operations but at the cost of complex mathematical operations, which could compromise its implementation in constrained environments. Rainbow, a multivariate polynomial signature scheme, offers a different approach but with larger keys, which could be problematic for certain applications in mobile networks. SIKE (Supersingular Isogeny Key Encapsulation) is a unique approach to cryptography based on the mathematical structure of elliptic curves. Due to its small key size and potential for efficient implementation, it could be used for user authentication and secure signaling and offers advantages in mobile applications where efficiency is key.


\textbf{In terms of data encryption and privacy preservation,}  public key encryption and key establishment algorithms such as NTRU and Kyber are essential for encrypting data and protecting privacy in mobile networks. NTRU is efficient and has been analyzed for a long time. It offers a good balance between security and performance in encrypting data and maintaining privacy. Kyber is known for its efficiency and smaller key sizes, making it suitable for encrypting data in mobile networks where bandwidth and storage space may be limited. BIKE (Bit Flipping Key Encapsulation) was developed as a \ac{KEM}  based on error correction codes \cite{aragon2022bike}. It is particularly suitable for establishing secure communication channels, which are the basis for encrypting data and maintaining privacy in mobile networks. Similar to BIKE, HQC (Hamming Quasi-Cyclic) is a KEM that relies on error-correcting codes \cite{melchor2018hamming}. It is designed to provide security for data encryption and privacy protection, with a focus on defense against quantum attacks.

Ensuring secure network management, infrastructure security, and efficient key management in mobile networks requires algorithms that are not only secure against quantum attacks but also efficient in terms of computation, bandwidth, and storage. \textbf{For network management and infrastructure security,}  algorithms such as Kyber, Classic McEliece, and NTRU can be used for secure channel establishment, which is essential for secure network management and infrastructure protection. Classic McEliece is based on code-based cryptography, which is known for its high level of security. It offers significant resistance to quantum computing attacks, making it a strong candidate for securing network infrastructures and encrypting data. The downside is that the key size is larger, making it better suited for static or less bandwidth-sensitive applications. SIKE's features make it a compelling option for secure network management, especially in environments with limited computing resources and bandwidth. \textbf{For key management,}  Crystals-KYBER, NTRU, SABER, and BIKE are considered efficient for key management tasks due to their balance between security and performance. Their efficiency in key encapsulation makes them suitable for dynamic and scalable key management in mobile networks. In addition, BIKE's efficient key encapsulation method could be advantageous for key management in mobile networks, as it enables the secure exchange of keys with relatively little computational overhead.

%% file: limitations.tex
\subsection{Limitations and Challenges}

Integrating the \ac{PQC} algorithms and QKD method into existing mobile network systems poses several technical challenges, including performance considerations, resource constraints, and interoperability issues. 


\subsubsection{QKD specific challenges}

\textbf{In terms of range and cost}, current QKD technology is prone to signal degradation at long distances, making it impractical for widespread use in large cellular networks. The technology is also currently expensive to implement and maintain. \textbf{In terms of infrastructure requirements}, QKD systems often require special equipment and dedicated fiber optic lines for secure transmission. This increases the complexity and cost of the network infrastructure. \textbf{In terms of integration into existing networks}, the seamless integration of QKD into existing mobile network architectures poses a technical challenge due to its unique characteristics and requirements. In addition, one of the biggest barriers to the widespread adoption of quantum technologies is the \textbf{general lack of awareness and expertise}, especially among companies in secondary markets and emerging economies. These organizations often lack an understanding of the potential threat that quantum computing poses to current cryptographic standards. In addition, the limited availability of quantum-safe technologies and solutions in secondary markets can lead to delays in adoption and adaptation due to high costs or prioritization of primary markets with advanced R\&D capabilities. To address this issue, it is important to provide cost-effective and open-source solutions so that international organizations can efficiently meet their security needs. The heterogeneity and complexity of global supply chains can also lead to vulnerabilities, especially if parts of the chain lag behind in adopting quantum security practices. This can jeopardize the entire system as components and services are sourced from different markets, including those that are considered second-tier or emerging.

\subsubsection{PQC specific challenges}

\textbf{In terms of computational overhead}, many PQC algorithms require more computational resources compared to classical cryptographic algorithms such as RSA and ECC. This increased computational overhead can lead to slower encryption and decryption speeds, which affects the overall performance of systems, especially in resource-constrained environments such as mobile devices or IoT devices. \textbf{In terms of key generation and management}, PQC algorithms often involve larger keys and more complex key generation processes, which can further exacerbate performance issues. Generating and managing these larger keys can require additional processing time and memory resources. \textbf{In terms of power consumption,} the increased computational requirements of PQC algorithms can lead to higher power consumption, which is particularly problematic for battery-powered devices such as smartphones or IoT sensors. A balance between security requirements and energy efficiency is important to avoid excessive battery consumption. \textbf{In terms of standardization and adoption,} PQC algorithms are still under development and there is an ongoing debate in the cryptographic community about which algorithms should be standardized and widely adopted. A lack of consensus on standardization can hinder interoperability between different systems and applications, leading to compatibility issues and fragmentation of the cryptographic ecosystem. \textbf{In terms of legacy system compatibility,} the integration of PQC algorithms into existing systems and protocols designed to use classical cryptographic algorithms can pose an interoperability challenge. Ensuring seamless compatibility between PQC and legacy systems requires careful consideration and may involve protocol upgrades or transition mechanisms.

\subsection{Opportunities}


The introduction of PQC also offers several potential opportunities to improve security and data protection in various areas. \textbf{In terms of resilience to quantum attacks}, by implementing PQC, organizations can harden their security infrastructure against the emerging threat of quantum computing and ensure the long-term confidentiality and integrity of sensitive information. In terms of enhanced security guarantees, PQC algorithms often provide more robust security guarantees than their classical counterparts. By using mathematical problems that are considered difficult even for quantum computers, PQC algorithms provide an additional layer of security that reduces the risk of cryptographic compromise due to vulnerabilities in the algorithm or advances in cryptanalysis techniques.

The introduction of PQC can improve \textbf{privacy protection} by ensuring the confidentiality of sensitive data transmitted over insecure channels. With the robust encryption provided by PQC algorithms, individuals can communicate securely without fear of eavesdropping or unauthorized access to their personal data, protecting their privacy rights. The proliferation of \textbf{Internet of Things (IoT) devices and embedded systems} has increased the attack surface for cyber threats. PQC algorithms optimized for resource-constrained environments offer the ability to protect these devices from potential quantum attacks and keep critical infrastructure, smart homes, and wearable technologies safe from breaches. The introduction of PQC can increase \textbf{trust and security} in digital transactions and communication. By using PQC algorithms, organizations can demonstrate their commitment to proactive security measures and data protection, fostering trust among customers, partners, and stakeholders in an increasingly connected and data-driven world. The widespread adoption of PQC algorithms can help establish \textbf{global security standards}, promoting interoperability and consistency of cryptographic practices across industries and jurisdictions. By using standardized PQC algorithms, organizations can ensure compatibility with international security frameworks and legal requirements.

\subsection{Recommendations}

\textbf{Hybrid design:} Applications that deal with the transition to quantum-safe cryptography are already on their way to the market even before its standardization is completed. Recently, Apple announced a new hybrid cryptography (classical (ECDH p-256)) + post-quantum key encapsulation (Kyber-1024)) for both initial key establishment and all rekeys afterward to secure iMessage\footnote{Online: https://security.apple.com/blog/imessage-pq3/, Available: February 2024.}. However, these keys are classically authenticated with ECDSA p-256 in order to sign with the keys generated in the secure enclave of the device. Messages are encrypted with AES-256 in counter (CTR) mode. This shows that the hybrid design approach is taken as a more suitable approach than to go with plan \ac{PQC}.

\noindent
\textbf{Collaborative efforts:} Overcoming the technical challenges requires the joint efforts of researchers, industry representatives, and standardization bodies. Developing efficient implementations of PQC algorithms that are optimized for specific use cases, improving memory and energy efficiency, and promoting consensus on standardization and interoperability are essential steps toward the widespread adoption of post-quantum cryptography in various applications and environments.